# Bending-like stress induced by solder joint under uniaxial tensile testing in 2G-HTS tapes: Impact and optimization approach

Dean Liu[1], Yue Wu[1,*], Haoliang Xiang[1], Xiaofen Li[1], Caida Fu[3,4], Chiheng Dong[3,4], Yue Zhao[1,2]

[1]School of Electrical Engineering, Shanghai Jiao Tong University, 200240, Shanghai, People's Republic of China

[2]State key Laboratory of High-Efficiency special cable technology, Shanghai Jiao Tong University, 200240, Shanghai, People's Republic of China

[3]Key Laboratory of Applied Superconductivity, Institute of Electrical Engineering, Chinese Academy of Sciences, 100190, Beijing, People's Republic of China

[4]University of Chinese Academy of Sciences, 100049, Beijing, People's Republic of China

**Abstract**

The reversible stress limit ($R_{\text{rev}}$) of second-generation high-temperature superconducting (2G-HTS) tapes is a critical performance indicator, typically characterized through uniaxial tensile testing. In practice, the accuracy of the measured $R_{\text{rev}}$ value is often compromised by stress concentration induced by the voltage tap solder joint. The present study investigates the underlying interference mechanism using integrated experimental and numerical methods. Mechanistic analysis reveals that under uniaxial tensile loading, the local geometric inhomogeneity introduced by the solder joint induces an external, bending-like stress in the vicinity of the solder joint, transitioning from additional tensile stress in the zone adjacent to the joint to additional compressive stress in the zone remote from it. When the solder joint is attached to the front surface of the tape (the side closer to the superconducting layer), the superconducting layer experiences localized additional tensile stress, triggering premature damage and early $I_c$ degradation. Consequently, an optimized back-surface soldering approach is proposed, which positions the superconducting layer in a localized compressive zone. Experimental validation demonstrates that the proposed approach effectively mitigates testing errors for various tape configurations. Notably, for the tape with a copper layer thickness of 5 μm, the measured $R_{\text{rev}}$ increased from 546 MPa to 734 MPa, corresponding to a 42% increase, and moved closer to the actual value. The findings provide essential insights for the precision characterization of the electromechanical performances (EMPs) of 2G-HTS tapes.

**Keywords**

REBCO tapes, uniaxial tensile testing, bending-like stress, solder joints.

## 1. Introduction

The second-generation high-temperature superconducting (2G-HTS) tape is pivotal functional material for cutting-edge technologies such as magnetic confinement fusion [1], magnetic levitation [2], superconducting generators [3], magnetic energy storage systems [4], and magnetic resonance imaging [5]. The tapes feature a complex, multi-layered composite structure consisting of a metallic substrate, ceramic buffer layers, a ceramic superconducting layer, and metallic stabilizers [6]. The core functional component is the $REBa_2Cu_3O_{7-x}$ (REBCO, where RE represents rare-earth elements) ceramic superconducting layer, which exhibits exceptional properties, including high critical current [7,8] ($I_c$), high critical magnetic fields [9], and high critical temperatures [10]. In most magnetic applications, 2G-HTS tapes (also called REBCO tapes) operate under extreme conditions, low temperatures, high magnetic fields, high currents, and must withstand complex mechanical loads generated by the synergy of thermal strain and Lorentz forces [11,12]. These loads induce a degradation in the current carrying capacity of the tape [13]. Due to the anisotropic nature of the tape structure, different loading directions, such as uniaxial tension [14], bending [15], delamination [16], and torsion [17], exhibit distinct mechanisms and degrees of impact on current carrying capability.

Uniaxial tension is one of the most common loading conditions, and its impact on the $I_c$ of REBCO tapes has been extensively researched [18-20]. In uniaxial tensile testing, the reversible stress limit ($R_{rev}$) is used to measure the ability to withstand tensile loads. Once the tensile load exceeds $R_{rev}$, the $I_c$ of the REBCO tape undergoes a sharp, irreversible degradation. International round-robin tests for uniaxial testing have shown that results sometimes exhibit significant intra- and inter-laboratory scatter, which primarily stems from variations in testing methods, the non-uniformity of $I_c$ along the length of the REBCO tape, and other factors related to the testing process [21]. During testing, voltage taps used to measure the $I_c$ are typically soldered to the front surface (defined in Tab. 1) of the tape, inevitably leaving behind solder joints. Research by Xiyang Su *et al.* revealed that during uniaxial tension, local shear stresses are generated near solder joints, causing premature damage to the REBCO tape at those locations and reducing the accuracy of the test results [22]. To circumvent the influence of solder joints, Zili Zhang *et al.* proposed a two-step procedure [23]. Although that procedure yields more accurate results, it does not allow for *in-situ* testing of the REBCO tape, involves complex operations, and requires relatively long samples, making it difficult to apply in practical testing scenarios. To identify a testing approach that mitigates solder joint interference while remaining simple to operate, this work focuses on the influence of voltage tap solder joints on uniaxial tensile testing. digital image correlation (DIC) was employed to characterize the axial strain field on the surface of the REBCO tape under axial loading. Furthermore, finite element method (FEM) was utilized to clarify the stress state of the REBCO tape near the solder joints during the loading process. Based on these findings, an optimization approach is proposed, soldering the voltage taps to the back surface of the REBCO tape, and the feasibility of this approach is experimentally verified.

## 2. Experimental details

This study utilized three different structures of REBCO tapes provided by SuperMag Technology (Shanghai) Co., Ltd., including two copper-plated tapes and one laminated tape, which specific parameters are detailed in Tab. 1. The uniaxial tensile testing was collaboratively completed using a fatigue testing machine and a four-probe testing platform. First, a short sample with a length of 6 cm was cut from the tape to be tested. A micrometer and a vernier caliper were used to precisely measure its thickness and width, respectively, in order to calculate the cross-sectional area for stress conversion. After soldering the voltage taps to specific locations on the sample surface, the sample was secured onto a dedicated fixture. The entire assembly was then mounted into the fatigue testing machine, as shown in the inset of Fig. 1 (a). Prior to starting the test, the entire fixture assembly was fully submerged in a liquid nitrogen environment. First, the initial critical current $I_{c0}$ of the sample was measured via the four-probe method, adopting a quench criterion of 1 μV/cm [24].

Subsequently, the axial tensile stress was applied in a stepwise, incremental manner. After each loading step reached the target stress, the load was held constant to perform the *in-situ* measurement of the loading critical current $I_c^{load}$. The stress was then completely unloaded to measure the unloaded critical current $I_c^{unload}$. After that, the tensile stress was increased to the next level, and the aforementioned measurement procedure was repeated. Generally, the determination of $R_{rev}$ falls into two categories based on either $I_c^{load}$ or $I_c^{unload}$. This study adopts $I_c^{unload}$ as the criterion, defined as follows: the curve of the normalized unloading critical current $I_{c,nor}^{unload}$ ($I_c^{unload}/I_{c0}$) as a function of the uniaxial tensile stress is fitted using the Boltzmann function; the stress value corresponding to the point where the first derivative of the fitted curve reaches -0.001 $MPa^{-1}$ is defined as the $R_{rev}$ of the sample.

Tab. 1 Tape specifications

| Designation | Tape I | Tape II | Tape III |
| --- | --- | --- | --- |
| Type | Cu stabilized tape | Cu stabilized tape | Cu laminated tape |
| Structure | (front surface) lamination/ Cu/Ag/REBCO/Buffer/C276/Ag/Cu /lamination (back surface) [25,26] | | |
| Stabilizer, thickness (μm) | Cu ~ 20 | Cu ~ 5 | Cu ~ 1 |
| Dimension, thickness（mm）× width (mm) | ~ 0.09 × 4 | ~ 0.06 × 4 | ~ 0.17 × 6 |
| External lamination | - | - | Cu (both side) (~ 50 μm) |

Following the uniaxial tensile tests, the voltage taps were removed from the samples via localized heating for subsequent performance characterization. The spatial distribution of $I_c$ across the sample was characterized using a Scanning Hall Probe Microscope (SHPM) equipped with a CH HALL 1AHD802FG Hall probe. Prior to testing, the sample was field-cooled to 77 K in an external magnetic field of approximately 0.3 T, which is higher than the full penetration field. The external field was then removed to measure the trapped magnetic field distribution. Upon completion of the SHPM testing, a high-speed tin-stripping solution was employed to chemically remove solder joint from the sample surfaces. During this chemical reaction, the solution was continuously stirred until the surface solder was completely dissolved. Subsequently, an aqueous $FeCl_3$ solution was utilized to remove the copper stabilizer, fully exposing the underlying silver stabilizer. Magneto-optical (MO) imaging was subsequently employed within an optical cryostat (Montana Instruments) to characterize defects in the sample with the exposed silver stabilizer. The sample was first zero-field-cooled to 50 K, after which the strength of an external perpendicular magnetic field was gradually increased while magnetic flux penetration images were captured in real time. *In-situ* DIC characterization via the XTDIC-CONST-5M system was performed directly on the pristine samples. Solder joints were attached to the front surface of the samples, and a high-contrast speckle pattern was sprayed onto the surface designated for observation. The tensile tests were conducted at room temperature under displacement-controlled loading, with a loading speed set to 0.1 mm/min and an image acquisition interval of 0.5 s. The FEM simulation was implemented based on the layered shell interface within the COMSOL Multiphysics software. The geometric model of the tape was simplified into a multi-layered composite structure consisting of a copper layer, substrate, buffer layer, superconducting layer, and another copper layer, with its two-dimensional planar projection shown in Fig. 3 (a). The tape dimensions were set to a width of 4 mm and a length of 8 mm. The solder joint was simplified as a cylinder with a diameter of 2 mm. The tensile stress was set to 600 MPa. Detailed parameters for each layer are provided in Tab. 2.

Tab. 2 FEM parameters

| Designation | Thickness (μm) | Elastic modulus (GPa) | Mesh elements |
|---|---|---|---|
| Substrate | 43 | 210 [27] | 20 |
| Buffer layer | 0.1 | 166 [28] | 5 |
| Superconducting layer | 1.2 | 157 [29] | 5 |
| Copper layer | 5/10/20/30/40 | 118 [27] | 3/6/12/18/24 |
| Solder | 100 | 27 [30] | 60 |

**3. Results and discussion**

Fig. 1 (a) illustrates the uniaxial tensile test results for a short sample of Tape I. In this test, the voltage tap solder joints were attached on the front surface of the sample. Based on the $R_{\mathrm{rev}}$ criterion defined in this study, the value was calculated to be 546 MPa. When the tensile stress remains below this threshold, the $I_{\mathrm{c,nor}}^{\mathrm{unload}}$ remains stable. Once the stress exceeds this value, $I_{\mathrm{c,nor}}^{\mathrm{unload}}$ exhibits a sharp declining trend. This behavior aligns closely with the physical definition of $R_{\mathrm{rev}}$, validating the rationality of the criterion. The tensile test was terminated when $I_{\mathrm{c,nor}}^{\mathrm{unload}}$ dropped to 0.55. The voltage taps were then removed via heating, leaving behind the solder residues, as shown in Fig. 1 (b). Subsequently, the sample was characterized using a Scanning Hall Probe Microscope (SHPM), and the distribution of the normal component of the trapped magnetic field ($B_z$) was mapped, as shown in Fig. 1 (c). The flow direction of the screening currents is indicated by the black dashed arrows. In regions far from the solder joints, the screening current flows along the length of the sample, whereas in the vicinity of the solder joints, a screening current redistribution phenomenon occurs along the transverse direction. Quantitative analysis based on the Bean critical state model equation ($\nabla \times \boldsymbol{B} = \mu_0 \boldsymbol{j}_c$) [31] reveals that the $I_c$ at the solder joints is approximately 50% of that in the remote regions, which matches the $I_{\mathrm{c,nor}}^{\mathrm{unload}}$ value at the end of the tensile test. This indicates that the damage is confined to the region immediately adjacent to the solder joint, with the superconducting layer elsewhere remaining largely intact. To further investigate, the solder joint and copper stabilizer on the sample surface were sequentially removed using tin stripper and $FeCl_3$ solution to expose the silver stabilizer. Magneto-optical (MO) characterization results were shown in Fig. 1 (d) ~ (g). As the applied field reached 75 Gs, magnetic flux penetrates the superconducting layer, with the sample edges exhibiting typical feather-like flux penetration zones [32]. At the positions corresponding to edges of the solder joint, two strip-like flux penetration zones extending along the sample width appeared, representing the defects formed in the damaged superconducting layer after loading. As the field increased, these two defects, measuring 1.6 mm and 2 mm in length, respectively, became clearly visible. The longer defect accounts for approximately 50% of the sample width, consistent with the transport $I_c$ test results after loading. These findings confirm that when voltage tap solder joints attached to the front surface of the sample, the superconducting layer near the joints suffers damage and develops defects prematurely during uniaxial tensile testing, leading to a localized degradation of the sample $I_c$.

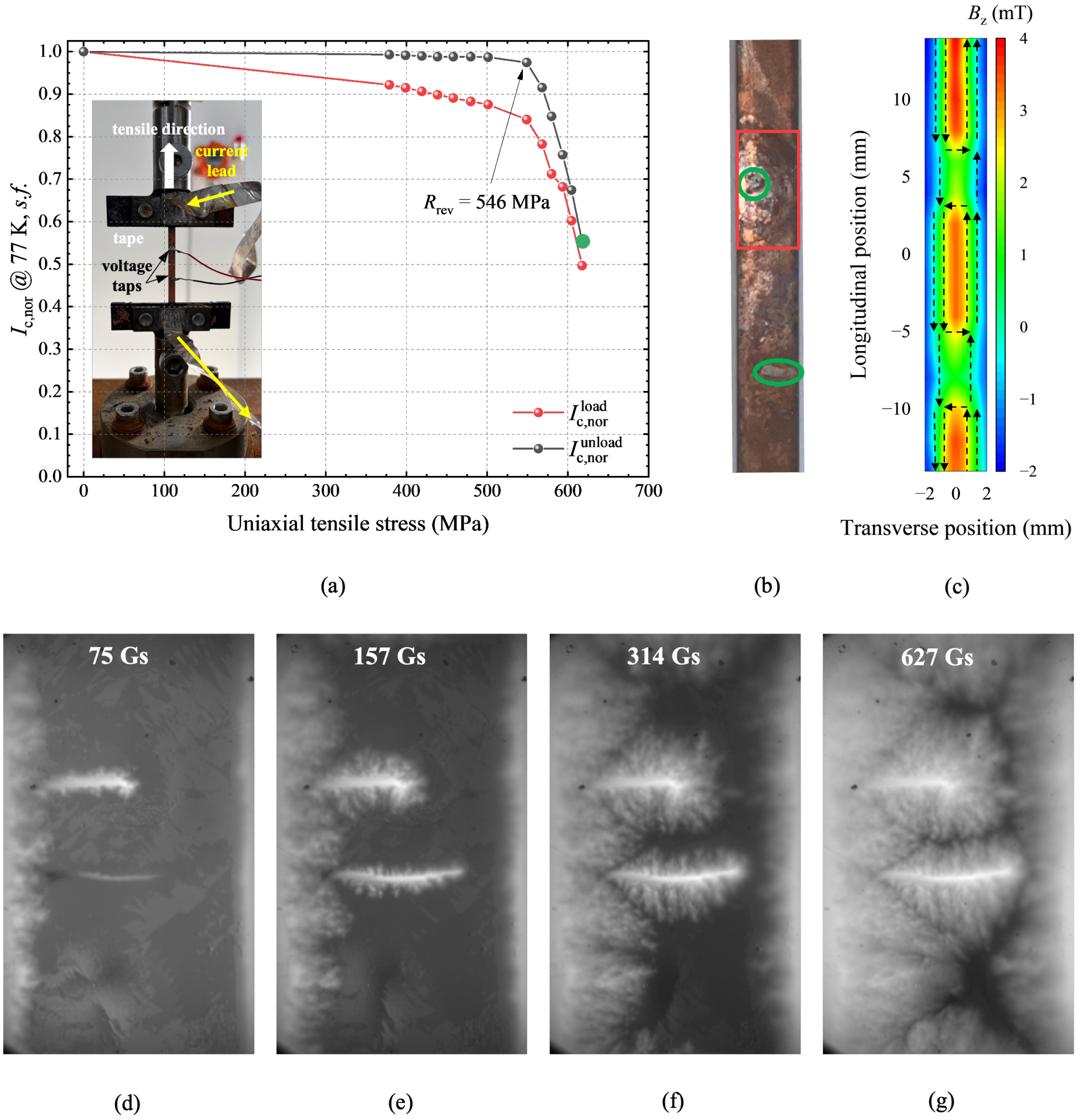

**Fig. 1** Uniaxial tensile test and post-mortem characterization of a Tape I short sample. (a) $I_c$ degradation as a function of tensile stress at 77 K and self-field (*s.f.*); the inset shows the experimental apparatus. $I_{c,nor}$ is $I_c/I_{c0}$ , $I_{c,nor}^{load}$ is $I_c^{load}/I_{c0}$, and $I_{c,nor}^{unload}$ $I_c^{unload}/I_{c0}$. (b) Optical photograph of the sample after the uniaxial tensile test with the voltage taps removed, and solder joints are highlighted by green circles. (c) SHPM results of the sample showing the normal component of the trapped magnetic field $B_z$; black dashed arrows indicate the induction current paths. MO images at 50 K under external fields of (d) 75 Gs, (e) 157 Gs, (f) 314 Gs, and (g) 627 Gs , respectively, focusing on the region within the red rectangle indicated in (b).

*In-situ* DIC characterization was performed on two short samples cut from Tape I during room-temperature tensile loading to analyze the strain distribution of REBCO tapes with solder joints attached to the front surface. For the first sample, the speckle pattern was sprayed on the front surface, as shown in Fig. 2 (a). The resulting axial strain fields under different loads are shown in Fig. 2 (b) ~ (d). Prominent high-strain regions consistently develop at both the upper-right and lower-left edges of the solder joint, with the strain concentration at the former being significantly more pronounced. This observation demonstrates excellent agreement with the MO characterization, where localized defects were found at the corresponding transverse edges of the joint. Simultaneously, the strain level within the solder joint region itself remains low due to the load-sharing effect. For the second sample, the speckle pattern was sprayed on the back surface, as shown in Fig. 2 (e), with the corresponding axial strain fields shown in Fig. 2 (f) ~ (h). During the tensile process, no obvious high-strain regions appeared on the back surface. Instead, only low-strain regions were observed within and adjacent to the projected solder joint region. This indicates that the stress concentration induced by the solder joint is not only spatially localized on the tape surface but may also exhibit a strong gradient along the thickness direction of the tape, which needs further investigation.

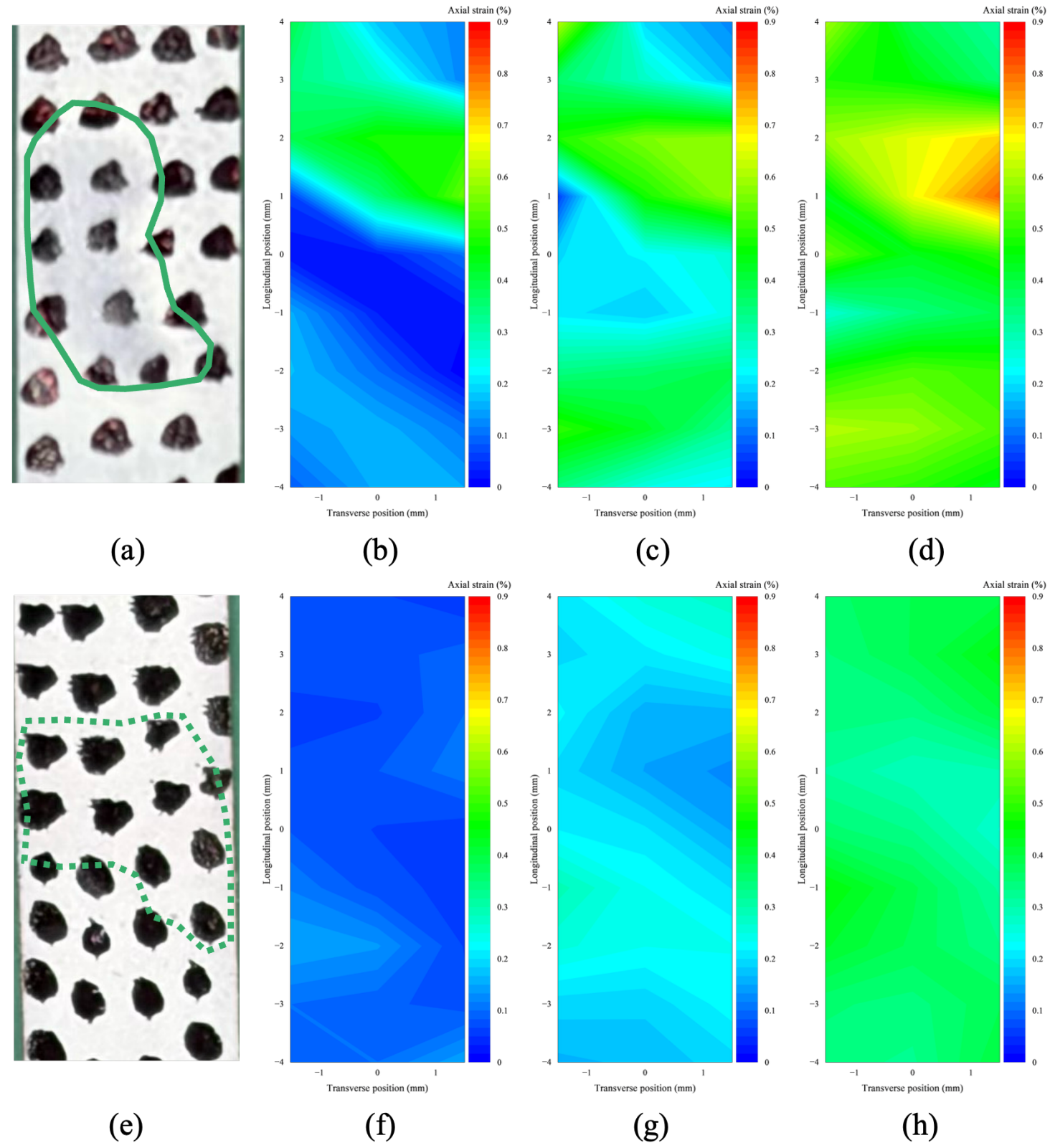

**Fig. 2** The DIC results of the Tape I. (a) Photograph of one sample front surface with speckle, and the solder joint on the front surface is circled by green irregular rectangle. Axial-field strain clouds under (b) 200 MPa, (c) 400 MPa, and (d) 600 MPa tensile stress. (e) Photograph of another sample back surface with speckle, and the solder joint on the front surface is circled by green dashed irregular rectangle. Axial-field strain clouds under (f) 200 MPa, (g) 400 MPa, and (h) 600 MPa tensile stress.

FEM was carried out to evaluate the stress distribution within the superconducting layer featuring the front-surface soldering joint. Firstly, the result of Tape I with 20 μm copper stabilizers was simulated, and the results are shown in Fig. 3 (b). The stress distribution of the entire superconducting layer was normalized using the stress below Point B, a region far from the solder joint that can be regarded as undisturbed by the joint. The normalized stress directly below the solder joint is significantly lower than in other regions, which is consistent with the low strain region identified in the DIC analysis. This is attributed to the load-sharing effect of the solder joint, where the solder bears a portion of the load, thereby relieving the stress on the underlying superconducting layer. Concurrently, significant stress concentration occurs at the transverse edges of the solder joint. This aligns perfectly with the two defects observed at the transverse edges in the MO characterization, as well as the upper and lower high strain regions captured in the front-surface DIC results. The blue curve in Fig. 3 (h) shows the stress variation from the front surface to the back surface below Point A, the location most heavily influenced by the solder joint. Because different layers possess distinct elastic modulus, the baseline stresses in the undisturbed regions vary. Therefore, the local stresses across different layers were normalized by their corresponding reference values below Point B. The normalized stress below Point A decreases monotonically from 1.14 at the front surface to 0.88 at the back surface. Regions with a normalized stress greater than 1 experience a local stress higher than the nominal applied load, whereas regions below 1 experience lower stress. This indicates that the impact of the solder joint is not a simple stress concentration, but rather a superimposed tensile stress near the joint and a compressive stress component further away. Concurrently, a neutral plane where the local stress equals the nominal load exists. This mechanical response is highly analogous to the stress state under bending, and is thus defined as bending-like stress. This finding is consistent with the bending deformation observed by Tianfa Liao *et al.* in simulations of solder-jointed tapes [33]. A prominent structural variation among different REBCO tapes is the thickness of the copper stabilizer layer. Thus, the stress distributions for tapes with varying copper thicknesses were simulated, as shown in Fig. 3 (c) ~ (f). For all configurations, high stress regions consistently appear below the transverse edges of the solder joint within the superconducting layer. As the copper layer thickness increases, the stress below Point A monotonically decreases, as summarized in Fig. 3 (g). Fig. 3 (h) illustrates the stress variation below Point A for different tapes. All configurations exhibit a similar bending-like stress profile to that of the 20 μm copper layer. Crucially, as the copper layer thickens, the maximum normalized stress monotonically decreases while the minimum value monotonically increases, demonstrating that the joint induced bending-like stress is progressively mitigated. These results reveal a potential strategy to mitigate the influence of the solder joint, positioning the superconducting layer near the neutral plane or as far as possible from the solder joint. Since the configuration of the tape is fixed, relocating the superconducting layer internally is impractical. Therefore, an optimized approach was proposed by placing the solder joint on the back surface of the tape.

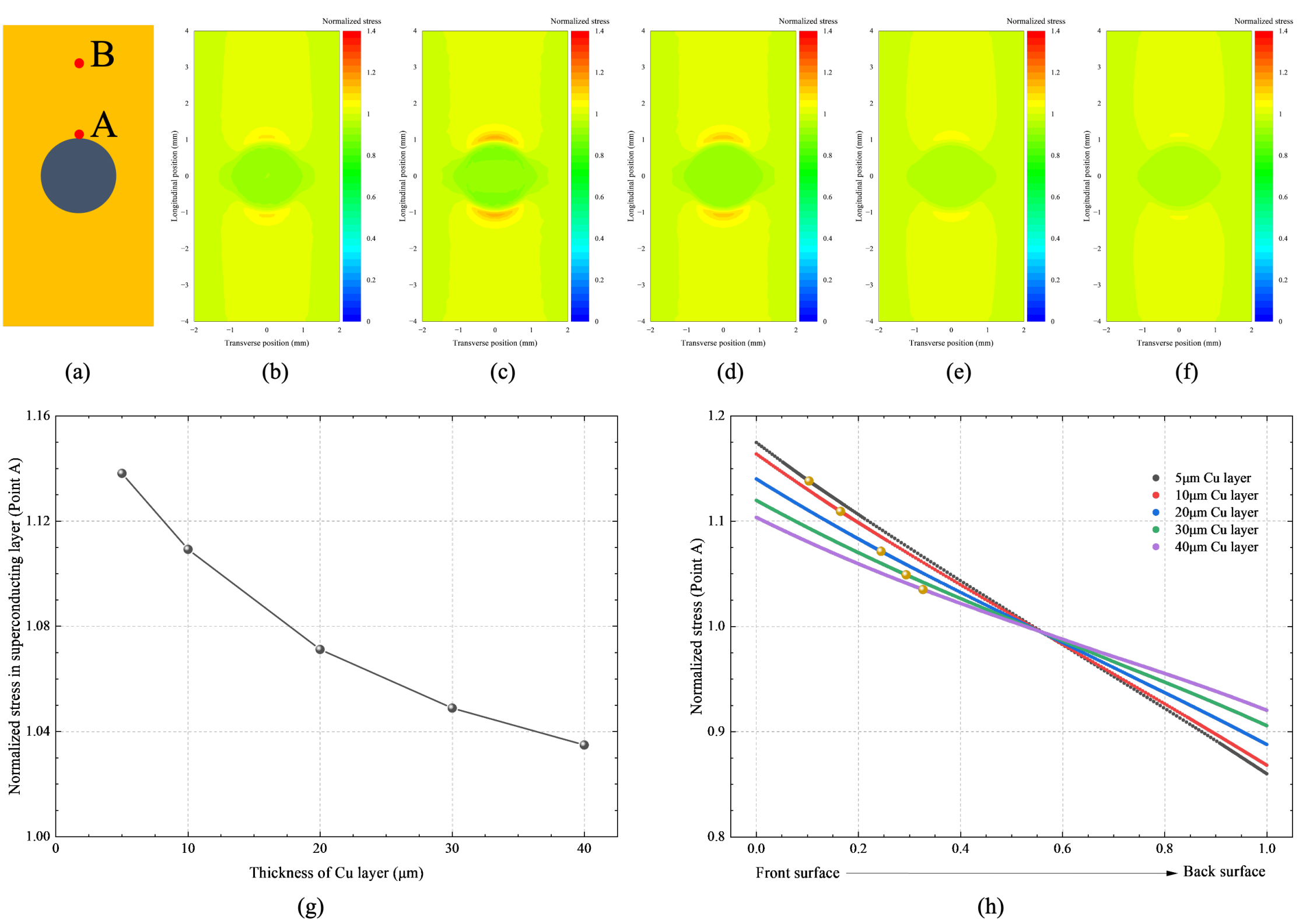

**Fig. 3** FEM results of REBCO tapes with solder joint attached to the front surface under 600 MPa uniaxial tensile stress. (a) Two-dimensional geometric structure of the simulation model, where the yellow rectangle represents the REBCO tape and the black circle represents the solder joint. Normalized stress distribution maps of the superconducting layer with copper layer thickness of (b) 20 μm, (c) 5 μm, (d) 10 μm, (e) 30 μm, and (f) 40 μm, normalized by the stress at red point B in (a). (g) Variation curve of the normalized stress below point A in the superconducting layer as a function of copper layer thickness. (h) Variation curves of the normalized stress below point A from the front surface to the back surface with various copper layer thickness. Yellow points represent the superconducting layer.

Fig. 4 presents the comparative results between the back surface soldering approach and the front surface soldering approach. Firstly, tests conducted on Tape I demonstrated that with the back surface soldering approach, the degradation of $I_c$ was significantly delayed, reaching an $R_{rev}$ of 734 MPa, an improvement of approximately 34% compared to the front surface approach. According to the FEM results, the superconducting layer far from the solder joint remains largely undisturbed, with its stress matching the macroscopically applied load. Therefore, the low-stress zone near the solder joint in the back surface approach does not lead to inflated test values, a higher tested value can be considered closer to the theoretical value. This indicates that the data obtained from the back surface soldering approach is more accurate. Furthermore, based on the FEM results for REBCO tapes with different copper layer thicknesses, a thicker copper layer weakens the influence of the solder joint. Correspondingly, the degree of improvement provided by the back surface approach should decrease as copper thickness increases. In this study, multiple sets of repeated experiments were performed on three types of tapes with varying copper thicknesses (Tape I, II, and III), as shown in the Fig. 4 inset. The $R_{rev}$ values for the three tapes improved by 36%, 42%, and 8%, respectively. Notably, Tape II with the thinnest copper layer showed the greatest improvement, while Tape III with the thickest copper layer showed the smallest, which is in high agreement with the FEM results. Statistical results confirm that by soldering the voltage taps to the back surface of the REBCO tape, the errors introduced by solder joints can be effectively mitigated, significantly enhancing the accuracy of uniaxial tensile test results. And this configuration proves to be universally applicable to REBCO tapes of various configurations.

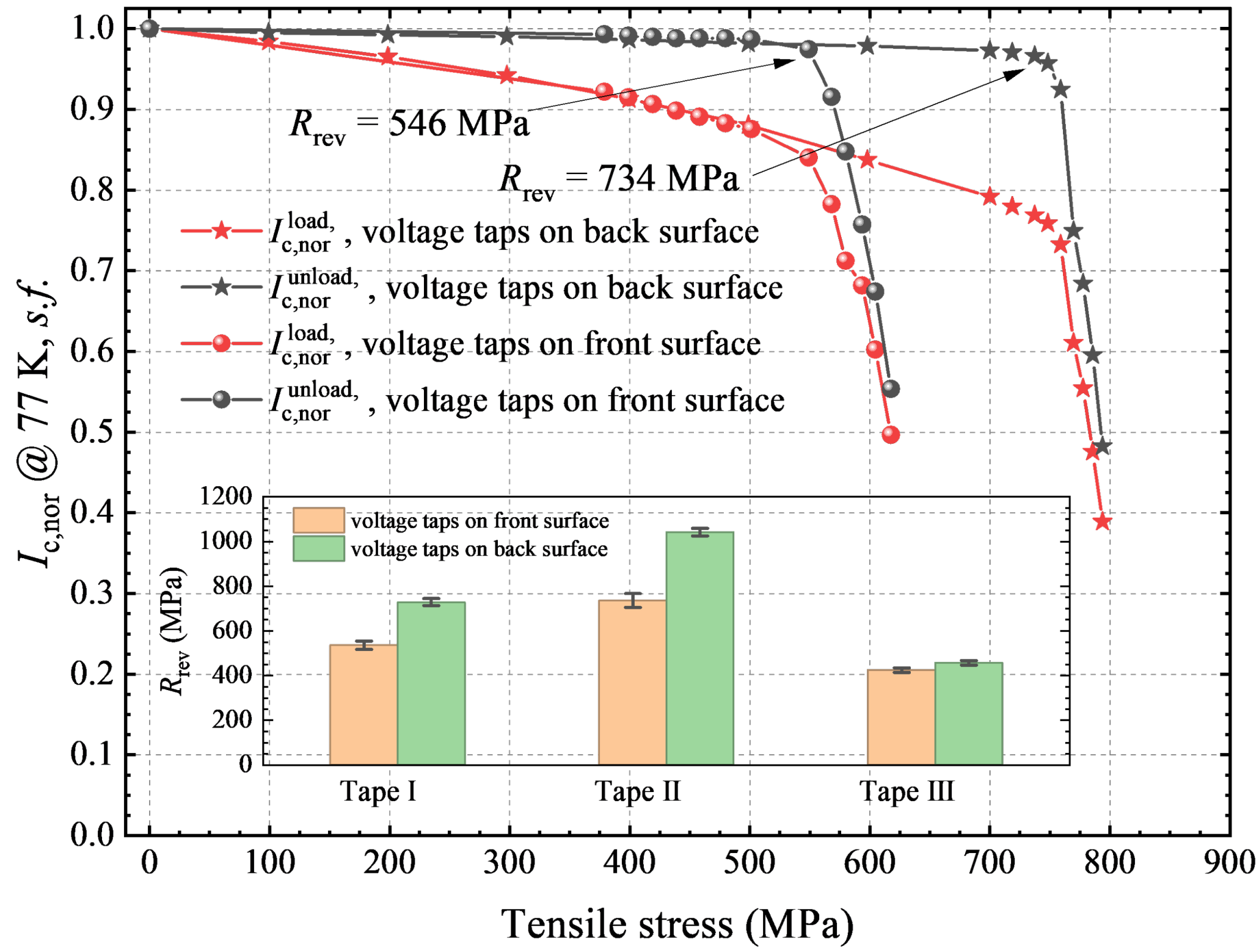

**Fig. 2** Uniaxial tension test results of Tape I with voltage taps attached to the front and back surface respectively. The inset presents a statistical comparison for Tape I, II, and III.

**4. Conclusion**

In summary, this study employs a combination of experimental characterization and numerical simulation to deeply explore the influence mechanism of voltage tap solder joints on the uniaxial tensile testing in REBCO tapes, proposes an effective optimization approach accordingly. It was found that REBCO tapes with voltage tap solder joints generate additional bending-like stress, which manifests as superimposed tensile stress on the side near the solder joint and as compressive stress on the side far from it. The degree of this additional stress gradually weakens as the copper stabilizer thickness increases. Since the superconducting layer is physically located closer to the front surface of the tape, the additional tensile stress results local defects and premature $I_c$ degradation in the superconducting layer, which in turn causes the tested $R_{rev}$ to deviate from the true value. Based on this mechanism, the back-surface soldering approach proposed in this paper effectively avoids the interference of localized high stress by moving the superconducting layer away from the solder joint. Experimental validation demonstrates that this approach mitigates the influence of bending-like stress on the tensile testing for REBCO tapes across various configurations. An increase in the measured $R_{rev}$ is consistently achieved for all tapes. Furthermore, the degree of this enhancement correlates inversely with the thickness of the copper stabilizer, perfectly consistent with the simulation results. This study not only reveals the fundamental nature of how soldering voltage taps affects the characterization of REBCO tape electromechanical performance but also provides an important scientific basis for establishing highly reliable testing methods for REBCO tapes.

**Acknowledgements**

This work was financially supported by the National Natural Science Foundation of China (Grant No. 52277027 and 52507030), National Key R&D Program of China (Grant No. 2024YFF0508003) and Science and Technology Commission of Shanghai Municipality (Grant No. 23511102003). The authors gratefully acknowledge Beijing Huahang Jingyi Technology Co., Ltd. for performing the DIC analysis and Dr. Jun Lu from the National High Magnetic Field Laboratory for his kind assistance with the arXiv endorsement.